\documentclass{article}

\usepackage{subfigure}
\usepackage{makecell}
\usepackage{stfloats}
\usepackage{PRIMEarxiv}

\usepackage[utf8]{inputenc} 
\usepackage[T1]{fontenc}    
\usepackage{hyperref}       
\usepackage{url}            
\usepackage{booktabs}       
\usepackage{amsfonts}       
\usepackage{nicefrac}       
\usepackage{microtype}      
\usepackage{lipsum}
\usepackage{fancyhdr}       
\usepackage{graphicx}       
\graphicspath{{media/}}     
\usepackage{amsmath}
\title{Application of Artificial Intelligence in Hand Gesture Recognition with Virtual Reality: Survey and Analysis of Hand Gesture Hardware Selection
}

\author{
  Jindi Wang  \\
  Durham University \\
  \texttt{jindi.wang@durham.ac.uk} \\
}

\begin{document}
\maketitle
  
\begin{abstract}
The ongoing usage of artificial intelligence technologies in virtual reality has led to a large number of researchers exploring immersive virtual reality interaction. Gesture controllers and head-mounted displays are the primary pieces of hardware used in virtual reality applications. This article analyzes the advantages and disadvantages of various hardware tools as well as possible use cases for hand gesture recognition. Several popular data sets used in data pre-processing and hand gesture detection systems are also summarized. This paper's primary objectives are to evaluate the current state of the art in gesture recognition research and to offer potential routes for future study to guide researchers working on immersive virtual reality interaction. Comparing recent study results and taking into account the state of the development of human-computer interaction technology, a number of potential research directions are highlighted, including the following: input data noise should be filtered, a lightweight network should be designed, tactile feedback should be created, and eye and gesture data should be combined.

\end{abstract}


\section{Introduction}
\label{sec:introduction}

Due to the recent rise of the concept of the ``meta-universe'', there has been a great deal of interest in human-computer interaction patterns in the meta-universe, which is essentially a virtual, digitised version of the real world. Also, this process of virtualization is like how a developer builds a virtual reality world, where hand gesture recognition is one of the most important technologies. The basic meaning of hand gesture recognition is that the computer can understand the user's hand gesture to promote human-computer interaction (HCI). Moreover, the human hand gesture is an essential source of information for human social interaction and plays a crucial role in human-computer interaction. In order to achieve natural and smooth interaction, computers need to understand human hand gestures and respond accordingly \cite{ji2019survey}. 

\begin{figure}[h]
  \centering
  \includegraphics[width=\textwidth]{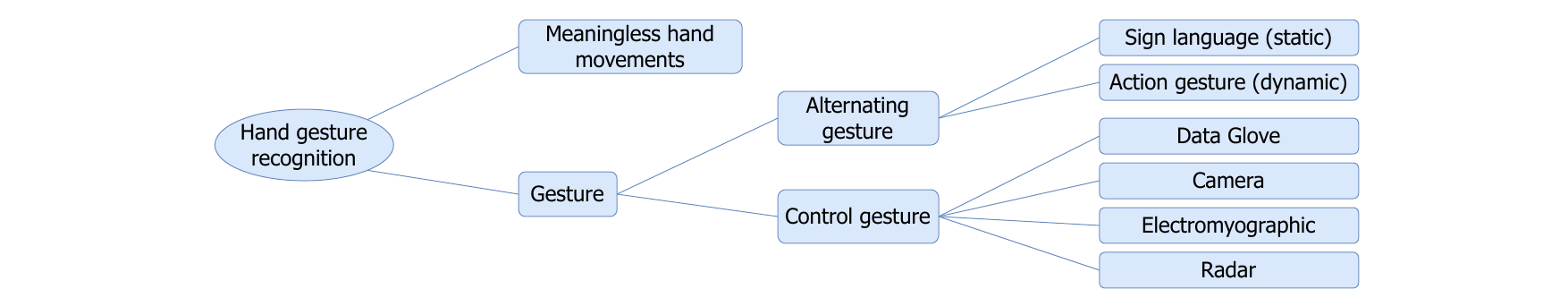}
  \caption{hand gesture recognition}\label{hand_gesture_classification}
\end{figure}

Additionally, hand gestures are a combination of a series of hand shapes over time, which can achieve specific tasks and express the user's thoughts and other functions \cite{jiang2020brief}. As shown in Fig.~\ref{hand_gesture_classification}, hand gesture recognition can be divided into meaningful and meaningless gestures. Meaningless gestures, such as the user shaking unintentionally and just tiny movements in the fingers, are primarily those that cannot accurately represent the user's ideas. Moreover, depending on the function, meaningful gestures can be divided into alternating gestures and control gestures. In contrast, alternating gestures can divide into static and dynamic gestures according to other temporal characteristics of gestures. Static gesture means that at a specific time, a gesture uses a particular shape to express a specific meaning, mainly including shape, direction and other features; Dynamic gesture refers to the combination of a series of gestures in a period of time; Control gestures could be classified according to different input devices, such as the four gesture recognition controllers mentioned earlier. In 2016, Clark \textit{et al.} \cite{clark2016system} made a report on the design, implementation and evaluation of a static gesture recognition system using Leap Motion Controller's virtual reality application and found the potential of a gesture recognition system based on machine learning to be applied in VR. Based on double-hand gestures in an interactive virtual environment, Reham \textit{et al.} \cite{rehman_two_2018} proposed a 3D navigation technology that allows users to control the speed during navigation effectively, and they use a gesture recognition engine to evaluate the accuracy and performance of the proposed poses. In 2020, the author \cite{kim2020vivr} proposed an immersive virtual reality technology to provide a realistic walking experience for people with visual impairment, conducted a survey comparing the experience with the real world, and found a high sense of presence in this virtual environment. Current virtual operation practices have limited functionality in guiding users to place objects in virtual environments. Nevertheless, Noghabaei \textit{et al.} proposed a snap-to-fit function to enhance the operation of the placement process by comparing two models \cite{noghabaei_object_2021}.

In the past two decades, the primary purpose of human gesture recognition has been to identify human movements or gestures in videos \cite{elmadany2018multimodal}, and this technology has been widely used in several research fields, such as healthcare \cite{aranki2014continuous}, security \cite{bloom2012g3d}, and human-computer interaction \cite{ren2013robust,nam2012intelligent}. A key component of a skilled human-computer interaction framework is gesture recognition and the use of signals as a feature interface that facilitates interaction \cite{sagayam2017hand}. As multimedia spreads, human beings are able to receive visual information more easily \cite{yang20183d}. Gesture can not only be used for voice commands, but also an essential component in HCI of Virtual Reality (VR) and augmented reality (AR) \cite{Ikram2020}. Also, it is essential in many applications including gaming, entertainment, media technology and continued advancements in virtual, augmented, and mixed reality (VAMR) technology \cite{wilk2018wearable}. 

In recent years, VR, Augmented Reality (AR), Artificial Intelligence (AI) and other technologies have been widely used in people's daily life. In order to give people a better sense of immersive experience, a large number of VR hardware devices have developed rapidly virtual reality interaction, and the essential virtual reality interactive devices are headset displays and gesture controllers. The most popular headsets for research used are HTC Vive \cite{Bickmann2019}, Oculus Quest, Oculus Rift, and Valve Index. At the same time, gesture controllers come in many different types and are used in different scenarios. Standard gesture controllers like based on data gloves \cite{meng_smart_2020}, based on electromyographic(EMG) signals \cite{zhang_framework_2011, nasri2020semg, zhang_novel_2020,wang2023exploring}, and based on radar sensors \cite{zhang_latern_2018, lee_improving_2020,wang2024impactarxiv} and based on cameras \cite{ikram_skeleton_2020,soton489244}. With VR products being more accessible \cite{kritikos2020comparison,li2024integrating,wang2024comparative}, it became possible for more specific VR applications to be created from a financial perspective \cite{blakey2016effects,bun2017low,wang2024impact}. Aditya \textit{et al.} \cite{aditya2018recent} introduced three emerging motion controller technologies, namely leap motion controller \cite{karashanov2016application,li2023towards,wang2023user}, Microsoft Kinect \cite{song2013real} and data glove \cite{fang20183d,chen2018design,wang2023developing}, and discussed the application, advantages and disadvantages of the three technologies.

With the hardware iteration of gesture input devices and the application of different types of devices in academic and industrial fields, researchers have to choose a suitable device for their own research. Nowadays, there are four main approaches to capture hand gestures such as glove-based devices, camera-based devices, radar-based devices and electromyography-based devices. Gesture recognition technology can be applied in all of these approaches, but there is no clear standard on how to choose an appropriate approach for research. Therefore, the main research question of this paper is \textit{\textbf{what factors will affect the researchers' choice of gesture recognition hardware equipment?}}

Summarising, the main contributions of this paper are described below:
\begin{enumerate}
    \item Summarise the current commonly used gesture recognition data acquisition methods, such as camera, data glove, radar sensor, Myoelectric ring; their advantages and disadvantages as well as suitable application scenarios are discussed.
    \item Summarise the commonly used gesture recognition data sets and data preprocessing methods.
    \item Summarise the commonly used deep learning models applied to gesture recognition.
    \item Explore and discover potential subdivision research directions in the current research field of gesture recognition.
    \item Provide a guideline for researchers who would like to work in the field of gesture recognition.
\end{enumerate}

The remaining components of this paper are described as follows. The Section \ref{sec:Method-PRISMA} mainly introduces the Preferred Reporting Items for Systematic Reviews and meta-Analyses (PRISMA) method used in this paper. Section \ref{sec: Hand Gesture Recognition and Input Hardware TODAY} mainly introduces the four hand gesture recognition system at present and summarise their advantages and disadvantages; The Section \ref{sec:Factors implication in hardware selection} introduces the factors implication in gesture hardware selection; the holistic importance of the identified factors to gesture input hardware selection is presented in Section \ref{sec: Holistic importance}. The Section \ref{sec:Data sets and pre-processing} mainly introduces the primary data set and discusses the advantages and disadvantages of using this data set. The Section \ref{sec: models} introduces several neural network models for gesture recognition and summarises the advantages and disadvantages of the models. The Section \ref{sec: Future Research agendas} introduces the current potential research direction based on the previous research results. The Section \ref{sec: Discussion} discusses the main research direction based on this survey in the future. Finally, we briefly conclude in Section \ref{sec: Conclusion}.


\section{Method-PRISMA}
\label{sec:Method-PRISMA}
\begin{figure}[h]
  \centering
  \includegraphics[width=15cm]{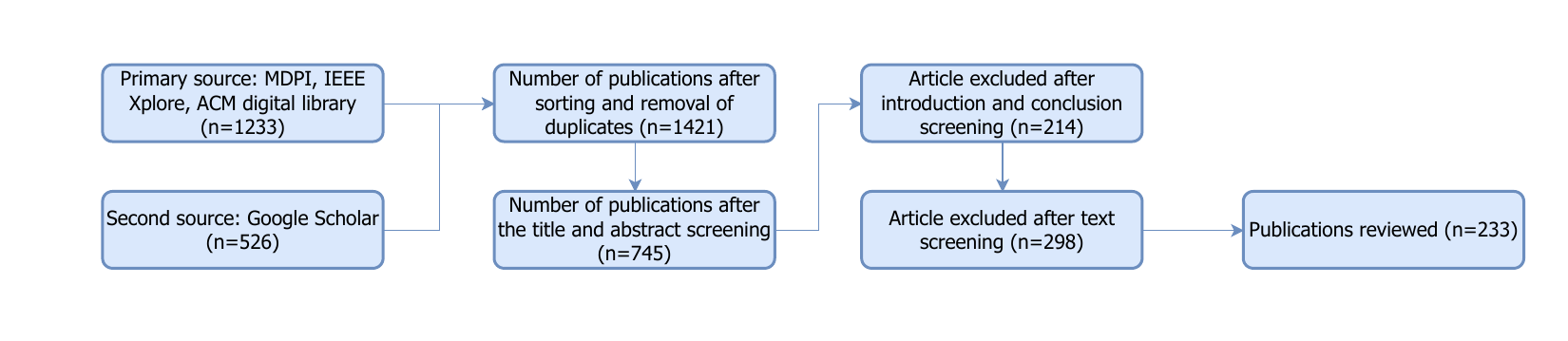}
  \caption{Main results of the PRISMA literature review stages}\label{fig: PRISMA}
\end{figure}

Preferred Reporting Items for Systematic Reviews and Meta-Analyses (PRISMA) is proposed by \cite{David_PRISMA_2009} to help researchers improve systematic reviews and meta-analyses, which mainly focus on randomized trials but can also be used as the basis for other types of systematic review studies. This paper investigates the application of gesture recognition in VR using PRISMA. The data are from published articles, as shown in Fig. \ref{fig: PRISMA}. The primary search databases are MDPI, IEEE Xplore and ACM Digital Library, and the secondary source is Google Scholar. Keywords: gesture recognition system, VR, immersive interactive system, data glove, Leap Motion, EMG signal, radar system, gesture control. Criteria are adopted and dropped by filtering titles and abstracts, followed by filtering introductions and summaries and then filtering full text. Although we found many references in multiple databases through our first screening, we focused on gesture recognition, and much literature was discarded as gesture recognition was not mentioned. Then there was a lot of literature thrown out, mainly because they only said gesture recognition and didn't apply it to VR scenarios.

Here we are mainly looking for a range from 2017 to 2021, and only literature written in English are included. The standards adopted in the literature specifically adopt gesture recognition technology (including different gesture input devices) and apply it to VR scenarios, so the reasons for excluded paper are as follows:
\begin{enumerate}
    \item Many articles mainly describe the current development of virtual reality technology and its application in particular fields and research on eye positioning, tracking and recognition. Still, they do not mention gesture recognition and gesture input equipment.
    
    \item Recently discontinued devices. Some articles use gesture input devices that have been discontinued. The most relevant cases are Cyberglove, CyberTouch, CyberGrasp and CyberForce, four different solutions proposed by Cyberglove System but ceased to be used in 2019. It isn't easy to buy this device in the market. So it has no significant effect on future research.
\end{enumerate}


\section{Hand Gesture Recognition and Input Hardware TODAY}
\label{sec: Hand Gesture Recognition and Input Hardware TODAY}
Human motion analysis is a promising research direction, which can be applied to virtual reality, bionic gestures, sign language translation, intelligent robots, video surveillance and other fields. At present, more and more researchers have begun to study immersive virtual reality, which interacts with the natural world through a highly immersive virtual environment. Based on the study, various applications are being developed, such as shopping, entertainment, etc. Still, these studies are dependent on physical devices and expensive, which leads to being difficult to be widely used. The use of gesture is a vital part of natural interaction, and the gesture data can be mainly divided into static and dynamic data. Static gestures, which are specific actions, are more straightforward than dynamic gestures, which are the rapid execution of a particular set of steps. Currently, gesture data can be divided into the following four types: infrared signal (radar-based gesture recognition system), image (camera-based gesture recognition system), gesture angular momentum and position information (glove-based gesture recognition system), and biological EMG signal (EMG-based gesture recognition system). Lee et al. compared VIVE Controller and data glove to find out which way is more suitable for interaction \cite{lee2017user}.

\subsection{HGR based on Radar System}
With the development of high resolution and small radar, human-computer interaction is constantly using new methods to try various radar devices to achieve better performance. However, the main development obstacle of radar-based gesture recognition devices is the same gesture movement, and the recognition accuracy of the model will also be reduced due to different movement modes. In order to solve this problem, \cite{lee2020improving} proposed a 60 GHz FMCW radar-based hand gesture recognition system consisting of two main parts, such as the domain discriminator, which through adversarial learning to achieve high recognition accuracy to minimize the differences between different people who produce the same type of gesture, the 3D-CNN feature extractor structure inspired by Inception Net and the parameters of extractor are effectively reduced, furthermore, the average classification accuracy in the offline and online test environment is 98.8\% and 90\% respectively. 

\begin{figure}[h]
  \centering
  \includegraphics[width=15cm]{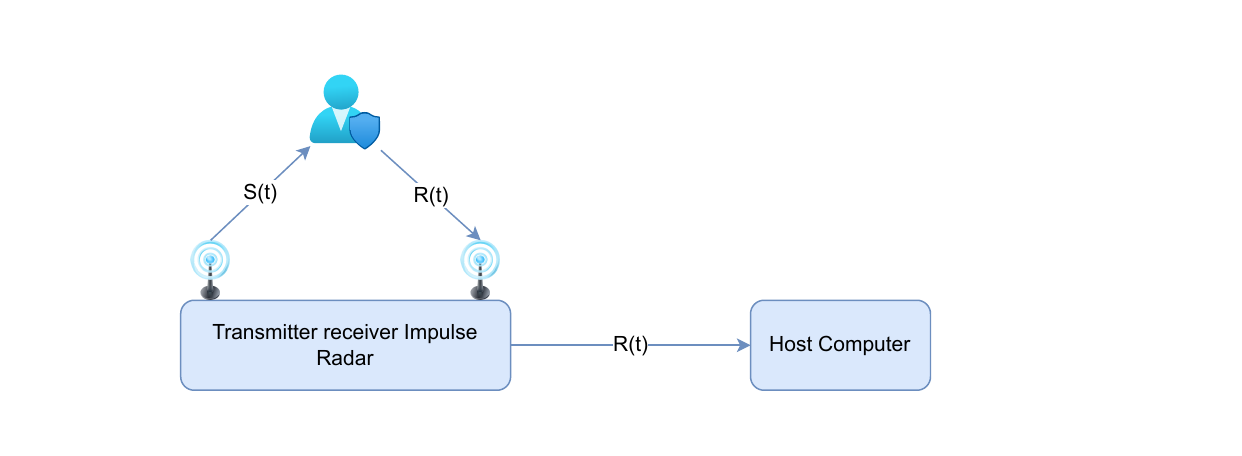}
  \caption{schematic diagram of radar signal transmission}
\end{figure}

With the wide application of gesture control in human-computer interaction, the performance of many radar-based systems will be reduced due to phase noise. When the system constantly receives the unprocessed radar data stream in the real-time environment, the known categories need to be detected synchronously to prevent noticeable delay. In order to solve such problems, \cite{zhang2018latern} proposed combining 3D-CNN and LSTM network as feature extractors and used the Connectionist Temporal Classification (CTC) algorithm \cite{zhang2018application} to classify the results with 96\% accuracy, compared with without using CTC the accuracy is 92\%. Since the signal of the object in front of the radar antenna can be very weak, \cite{zhang2018latern} tested different signal amplifiers at different test distances and found that the recognition accuracy was relatively stable within 2m. Still, the accuracy would decline after exceeding 2m due to the low signal-noise ratio.

Dynamic hand gesture recognition using the microwave or millimetre-wave radar sensors has become a standard technology for many human-computer interaction (HCI) applications; a novel method is proposed for dynamic hand gesture recognition based on micro-Doppler radar signatures by \cite{wang2020dynamic}. The short-time Fourier transform is carried out on the raw data to obtain the time-frequency spectrogram. The time-frequency spectrograms are associated with the same dynamic hand gesture model by a hidden Gauss-Markov model (HGMM), and the maximum likelihood criterion recognizes the testing gesture. Experimental results with actual radar data demonstrate that the proposed method has a strong generalization ability for radar gesture recognition in the cases of low signal-to-noise ratio (SNR) and unknown users.

\subsection{HGR based on Camera System}
\begin{figure}[h]
  \centering
  \includegraphics[width=15cm]{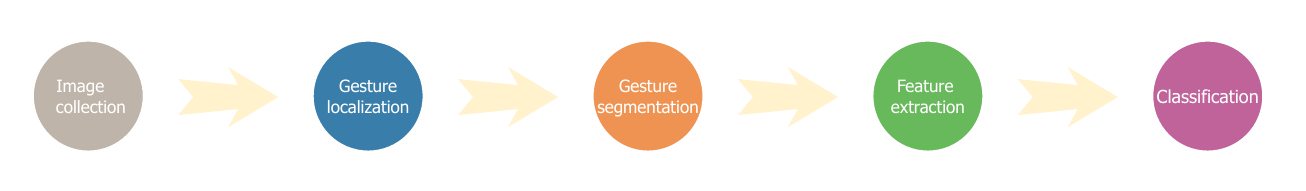}
  \caption{hand gesture recognition workflow of camera system}\label{workflow of camera system}
\end{figure}

The most popular data structure in gesture recognition is based on the image. Figure 4 shows a person using a camera to interact with the computer and generate images of the human gesture, gesture recognition system through image data collection, gesture segmentation, feature extraction, model training, and prediction to produce feedback, as shown in Fig. \ref{workflow of camera system}.

Although some researchers began to study to eliminate the dependence on physical devices of virtual reality interaction, these solutions are not entirely solving the reliance on the equipment or the environment. To solve this problem, Kang \textit{et al.} proposed the DeephandsVR \cite{kang2020deephandsvr} interface to use the CNN model based on the structure and parameters of Inception V3 to predict gestures, which only requires the hand to interact without other additional devices. They applied the model to an arcade game and conducted classification tests on six gesture categories. The result shows that the highest and lowest accuracy reached 94.7\% and 73.35\%, respectively. After analysing the impact that the low accuracy of classes may be due to a lack of training data, the data can be further added to solve the problem of insufficient data, as well as due to the lack of target categories; the applicable scenarios are limited, so the variety of object classification can be further expanded. Placidi \textit{et al.} proposed a virtual glove (VG), whose essence is to provide detailed 4D real-time hand tracking by using two orthogonal LEAP motion sensors \cite{placidi_data_2021}. VG has many applications in human-computer interaction, such as remote control of machines or remote rehabilitation \cite{Liu2015_remotecontrol, Mizera2020,Wang2020,wangchi2024,li2023broader}. For VG, a data integration strategy based on speed calculation is proposed \cite{placidi_data_2021}.

Biometric recognition based on image information has been an important field of computer vision research in recent years. Related research results are primarily used in natural human-computer interaction, virtual reality, intelligent video surveillance and other areas. In addition, the study of human gesture recognition is an essential part of human biometric recognition. However, due to individual differences, complex deformation, spatial and temporal changes of gestures, and visual amblyopia, these difficulties and reasons make gesture recognition a very challenging research field. Therefore, Li \textit{et al.} proposed a multi-sensor information fusion model \cite{li2019gesture} for gesture recognition virtual environment, and the method's effectiveness was verified through experiments. The results show that the recognition success rate of the proposed multi-sensor information fusion model can reach 96.17\% in the interactive virtual environment. Nonetheless, due to the real-time demand in the virtual reality environment, the scale of the neural network is not large, which limits the accuracy of feature segmentation. As a result, this method cannot solve the problem of multiple gestures' problem with the same starting or ending features and more complex issues. The advantage of this paper is that the problem of traditional gesture segmentation can be easily solved by data correlation processing in the information fusion model and dynamic sliding window, which makes gesture segmentation very simple.

Some authors \cite{lim2020video} studied a video-based gesture recognition method, which combined three-dimensional hand features, extracted spatio-temporal data, and carried out two-dimensional gesture classification under the support vector machine model. This method has good performance in static gesture recognition methods. Chanu \textit{et al.} \cite{Chanu2017}compared two different vision-based gesture recognition techniques with a gesture recognition technique based on data gloves and found that the vision-based approach was more stable than the glove-based technique under white light conditions.

\subsection{HGR based on Data Glove}
With the rapid development of human-computer interaction, data gloves have become an essential part of human-computer interaction. More and more companies research and develop data gloves. Now commonly used commercial data gloves can be divided into exoskeleton gloves and fabric gloves. The exoskeleton, which refers to a structure on the back of the hand, consists of strings or rigid links attached to the finger that provide feedback to the hand; The fabric, which refers to a piece of cloth covering the entire hand and fingers, includes many sensors and actuators to perform the desired function within the fabric, as shown in Figure 6. 

\begin{figure}[h]
  \centering
  \includegraphics[width=12cm]{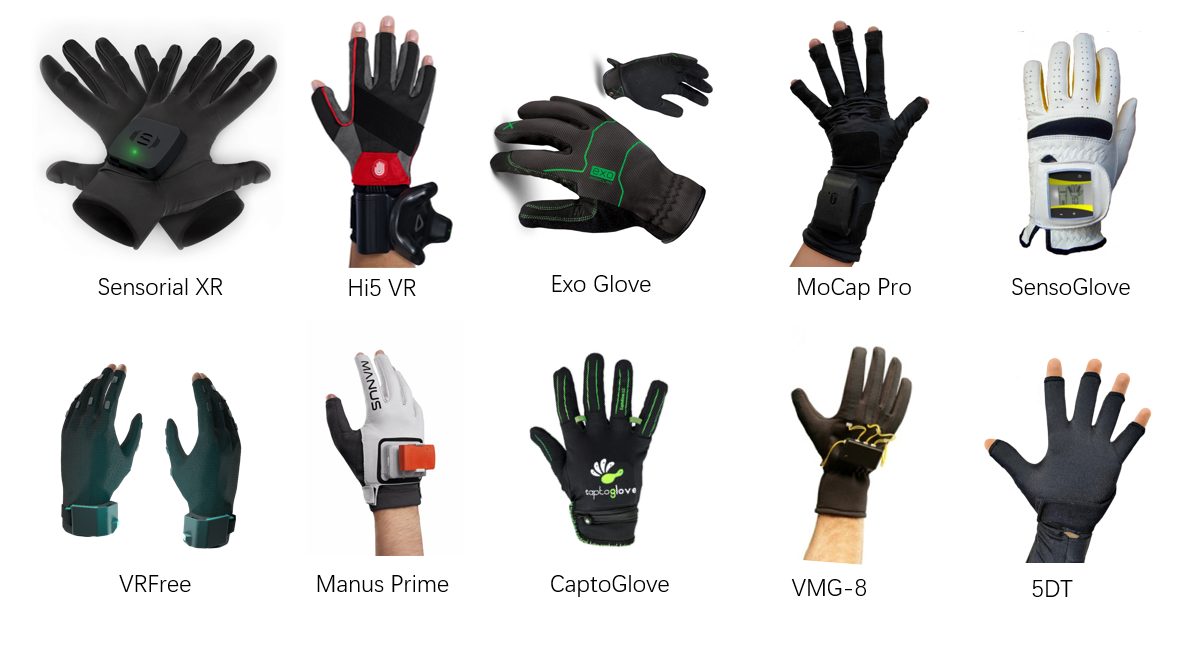}
  \caption{commercial data gloves}
\end{figure}

Data gloves have many applications, including enabling new types of human-computer interaction and automatically recognizing large gestures, such as those used in sign language. What is vital for most of these applications is to build mobile applications that do not require frequent communication with other services on the back-end server. Mummadi \textit{et al.} proposed a prototype of a data glove based on multiple small inertial measurement units (IMU) \cite{rahim2018human}. They collected a data set of 57 people using a glove-embedded French sign language classifier to capture the changes of different gestures when they used sign language to spell letters. The accelerometer on the sleeve is in a continuous slight fluctuation due to the involuntary shaking of the hand. Unlike the accelerometer, it is not affected by any external forces, but gyroscopes tend to drift from long-term averages due to the lack of a reference frame. Therefore, the solution proposes the fusion of sensor data from different sources to reduce uncertainty and improve the quality of a single series of readings. In this case, the accelerometer and gyroscope data were fused to eliminate problems caused by noise and drift, respectively. They used the random forest as the classifier was 92\% average accuracy and 91\% F1 score. The advantages of this model are the use of 5 IMUs and a multiplexer. The number of components is minimized to make glove-wearing more comfortable. By merging the data, complementary filters generate a smooth and consistent signal to obtain accurate finger direction and direction of movement. However, its disadvantage is poor performance in the real-time test. Wilcocks \textit{et al.} \cite{Wilcocks2021} has proposed a novel, consumer-grade data glove that provides precise user interaction and gives users a more immersive experience by employing simulated haptic feedback.

Since most of the current gesture recognition research is based on the data captured by the camera, which is often subject to the interference of the environment and is highly uncertain, most of the study based on the data glove adopts a large number of sensors to capture the data to achieve higher accuracy, which leads to the increase of the cost. In order to protect data collection from the influence of the external environment and reduce the cost of gloves, Giin \textit{et al.} \cite{giin2017smart} proposed a wearable gesture recognition system based on the movement of hands and fingers and to minimise the use of sensors; the author only used three IMU sensors to wear on the thumb, index finger and back of hand respectively. The performance of dynamic time warping was evaluated using a simple data set classified by six gestures, and the final test achieved an average accuracy of 93.19\%.

Since raw data glove data is very redundant. For example, the ``thumb up'' gesture exercises only the thumb joint, leaving the rest of the hand relatively at rest. So the author \cite{lv2017sparse} takes the lead in using the sparse decomposition model, which decomposes the data of data glove into two crucial parts (dictionary part and weights), then the dictionary part using support vector machine classifier achieve recognition. As a result, the average accuracy rate can reach 93.19\%, compared with the traditional KNN classifier, average accuracy is 78\% had the apparent promotion.

In 2019, Diliberti \textit{et al.} \cite{diliberti2019real} designed a pair of motion-capture gloves and applied them to obtain 3D rotation data of finger joints. At the same time, a network reduction strategy was proposed to appropriately reduce the complexity of network depth and width dimensions while ensuring high accuracy.

Although a lot of valuable research has been done on gesture recognition, there is still a lack of fine-grained gestures that can be effectively captured and tracked long-distance dependency in a complex environment. Hence, the author \cite{yuan2020hand} proposed and designed a novel data glove to capture gesture data. Meanwhile, in order to track the feature changes of gesture data, the convolutional neural network based on the feature fusion strategy is used to fuse the data collected by multiple sensors. In addition, the residual module is used to prevent over-fitting and gradient vanishing when deepening the network. Then the long short term memory network (LSTM) with fused feature data is used as the input of the classifier. Finally, the author through the two data sets (American sign language and Chinese sign language dataset data sets) for performance comparison of those of the other traditional neural networks. The experiment found that the proposed method can provide effective classification, especially in the American sign language data set can reach 99.93\% accuracy; the Chinese sign language data set, whose accuracy can only get 96.1\% because of its higher complexity, compared with the traditional model, it has noticeable improvement. Gunawardane \textit{et al.} \cite{gunawardane2017comparison} developed a data glove and compared its performance with Leap Motion Controller, and found that the average percentage error of the bending angle of Leap Motion and Data Glove on a single soft finger was 26.36\% and 18.21\%, respectively.

Gesture recognition provides an intelligent, natural and convenient way for human-computer interaction. Fang \textit{et al.} \cite{FANG2018198} proposed a gesture capture and recognition data glove based on the Inertial and magnetic measurement unit (IMMUs) and proposed static and dynamic gesture recognition methods based on extreme learning. An intelligent glove based on an inertial measurement unit (IMU) was developed by Meng \textit{et al.} in 2020 \cite{meng2020smart}. Based on visual, gesture, and speech data, they developed an intention perception algorithm and demonstrated that it could greatly improve the human-computer interaction experience. Zhang \textit{et al.} \cite{Zhang2020} proposed a real-time static and dynamic human gestures capture and recognition method based on data gloves, using a radial basis function neural network (RBFNN). The solution can process both static and dynamic gestures.In 2021, Dong \textit{et al.} \cite{Dong2021} designed a low-cost and efficient data glove. They proposed a new dynamic gesture recognition algorithm based on a convolutional neural network and time-domain convolutional neural network to recognize human dynamic sign language as well as mined and classified the long and short time dependence of gesture features. Based on a lot of experiments, it is found that the performance of this algorithm is better than many new algorithms.

\subsection{HGR based on Myoelectricity sign}

VR is a highly developed field, gesture tracking and recognition are two key technologies in mobile VR applications \cite{vulgari2019hand}. In the medical area, some patients cannot normal activities accord with human body kinematics of gloves and other equipment for them is not easy to use, hence, in order to make this part of the patients can feel a normal way of life, there is a lot of research using electromyography signals and virtual reality technology to meet the demand of patients and can fully understand the contents of the patients have to express. The authors proposed a gesture recognition framework based on a three-axis accelerometer and multi-channel EMG sensor information fusion and adopted a decision tree and multi-flow hidden Markov model as decision fusion, this approach can promote intelligent and natural control of gesture interaction \cite{zhang2011framework,li2023deep,li2023sim}.

\begin{figure}[h]
  \centering
  \includegraphics[width=12cm]{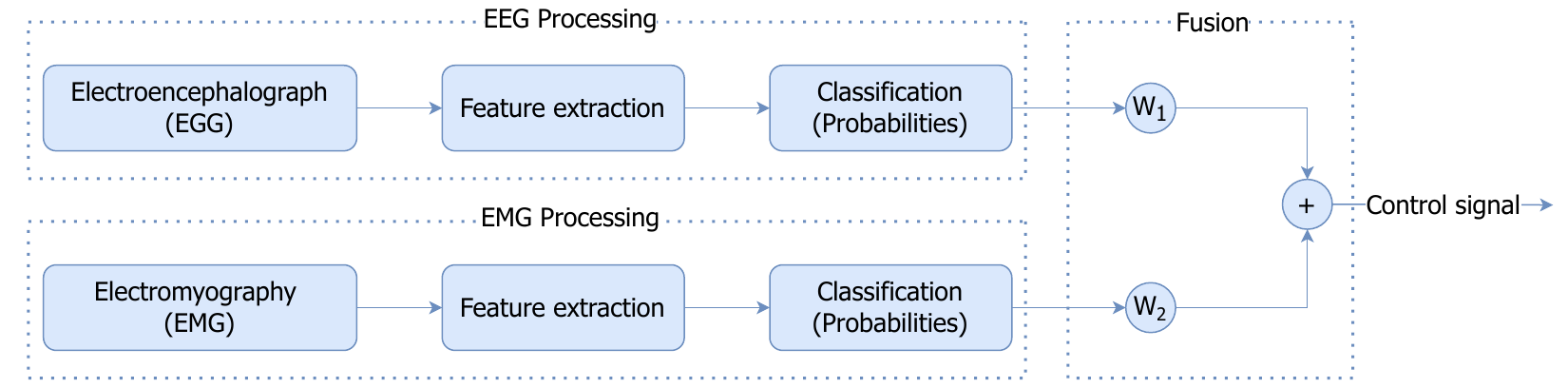}
  \caption{Schematic diagram of the combination of EMG and computer-brain interface}
\end{figure}

In 2017, Yang \textit{et al.} \cite{yang2017semg} proposed a recognition frame of continuous hand gestures to detect the person's intention for the control of the upper-limb exoskeleton robot. This frame is mainly concentrated on dynamic segmentation and real-time gesture recognition based on sEMG. The hand gesture was modelled and decomposed using Gaussian Mixture Model-Hidden Markov Models (GMM-HMM). GMMs are employed as sub-states of HMMs to decode the sEMG feature of motion. The log-likelihood and KL-divergence threshold are adopted to select the target gesture model. In myoelectric control schemes, the sEMG data are collected by Myo armband 8-channels sEMG sensors. The proposed framework has ideal classification accuracy, and its simpler acquisition armband makes it attractive to a real-time myoelectric control system.

A method based on a support vector machine (SVM) presented \cite{chen2019hand,li2021survey} to recognize hand gestures using surface electromyography (sEMG). In this method, the Myo armband (an EMG device with eight channels) measures subjects' forearm sEMG signals. The original EMG signal was preprocessed to reduce noise and detect muscle activity regions. Feature extraction is applied by segmenting a sliding sub-window in the preprocessed signals to get each segment of signals. Connect the signal segment with the results using a bag of functions to generate a feature vector. We train an SVM classification model for classification, which includes five sub-models. Each sub-model can recognize a gesture, like a fist, wave in, wave out, fingers spread, and double pinch. Finally, the authors test the proposed model to recognize these gestures and achieve an accuracy of 89.0\%.

In previous studies, various machine learning methods have been used to classify human actions, such as the K-nearest neighbour algorithm, Gaussian mixture model, support vector machine, Hidden Markov model, random forest and other technologies. Although the performance of the previous methods is good, the complex process of information extraction will inevitably lose helpful information. In order to solve this problem, Zhang \textit{et al.} \cite{zhang2020novel} proposed a new method using the RNN model that can give instantaneous prediction results. The raw surface electromyographic signal (EMG) data set containing 21 target classifications of 13 subjects were used to train the model, which gives a significant result that the model can predict the outcome before gestures to complete when the time step is 200 ms, and the accuracy was 89.6\%. However, we can do further research to accelerate the model's response speed further, improve the ongoing learning ability of the model, and improve the model structure, such as building a system similar to a generative adversarial network to ensure the safety of the model.

An effective transfer learning (TL) strategy was presented by Chen \textit{et al.} \cite{chen2020hand} for the realization of surface electromyography (sEMG)-based gesture recognition with high generalization and low training burden. To realize the idea of taking a well-trained model as the feature extractor of the target networks, 30 hand gestures involving various states of finger joints, elbow joints and wrist joints are selected to compose the source task. A convolutional neural network (CNN)-based source network is designed and trained as the general gesture EMG feature extraction network. Then, two types of target networks, in the forms of CNN-only and CNN+LSTM (long short-term memory), respectively, are designed with the same CNN architecture as the feature extraction network. Finally, gesture recognition experiments on three different target gesture datasets are carried out under TL and Non-TL strategies, respectively. The experimental results verify the validity of the proposed TL strategy in improving hand gesture recognition accuracy and reducing the training burden. For both the CNN-only and the CNN+LSTM target networks, on the three target datasets from new users, new gestures and different collection schemes, the proposed TL strategy improves the recognition accuracy by 10\% ~ 38\%, reduces the training time to tens of times, and guarantees the recognition accuracy of more than 90\% when only two repetitions of each gesture are used to fine-tune the parameters of target networks. The proposed TL strategy by \cite{chen2020hand} has essential application value for promoting the development of myoelectric control systems.


\section{Factors implication in hardware selection}
\label{sec:Factors implication in hardware selection}
There has been extensive research on machine learning technologies that use camera-based devices for gesture recognition, such as Leap Motion. Still, there has been limited research on the application of machine learning technologies in VR. PCVR devices like HTC Vive and Oculus Rift use hand-held remote interaction devices. These devices can accurately track the hand's position in the VR world but rely on buttons rather than gestures to some extent in VR interaction. In the future, VR systems are expected to use natural gestures as the primary input, so it is necessary to explore different methods to capture gesture data. From different perspectives, the four factors of hand gesture hardware selection are as shown in Fig~\ref{fig:factors}: 

\begin{figure}[h]
    \centering
    \includegraphics[width=0.8\textwidth]{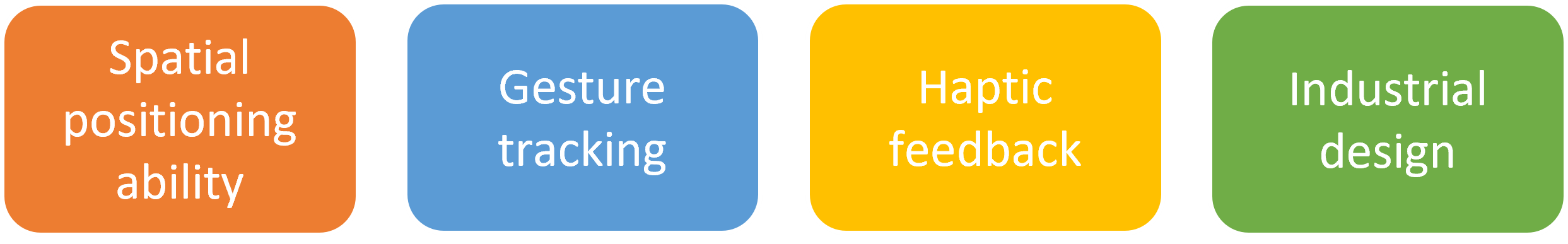}
    \caption{Four factors of hand gesture hardware selection}
    \label{fig:factors}
\end{figure}

\begin{table}[h!]
    \centering
    \caption{Four factors of hand gesture input hardware selection.}
    \label{tab:factors}
    \begin{tabular}{|p{3.5cm}|p{8cm}|p{2cm}|}
        \hline
        \textbf{Factors} & \textbf{Definition} \\
        \hline
        Spatial positioning ability &  obtaining the exact relative position of hand and human eye   \\
        \hline
        Gesture tracking & accurate, reliable and low-delay tracking   \\
        \hline
        Haptic feedback & accurately gives feedback on the touch sensation of objects touched by users   \\
        \hline
        Industrial design & physical peripherals should be lightweight to reduce the weariness of users, without affecting the movement of users' hands, and ergonomic comfort should be worn  \\
        \hline
    \end{tabular}
\end{table}

\subsection{Spatial positioning ability}
\textbf{Spatial positioning ability} is critical in hand gesture recognition systems as it determines the relative position of the hand to other objects or interactive surfaces within the system. Effective spatial positioning relies on accurate sensors and sophisticated algorithms capable of differentiating the hand's orientation and movement in a three-dimensional space. Technologies such as infrared sensors, stereo vision cameras, and depth-sensing technologies like those employed by the Microsoft Kinect or Leap Motion provide the backbone for capturing precise spatial data. These technologies allow systems to accurately detect hand positions, improving the system’s ability to interpret gestures as intended by the user.

\subsection{Gesture tracking}
\textbf{Gesture tracking} involves continuous monitoring and interpretation of hand movements. For a system to effectively track gestures, it needs to accurately capture a wide range of motions at high speeds with minimal latency. This requires robust image processing algorithms and possibly the integration of machine learning techniques to improve the recognition accuracy over time. Gesture tracking technologies should be capable of distinguishing between complex gesture patterns including both static (e.g., hand signs) and dynamic (e.g., waving or swiping) gestures. Advanced tracking systems utilize a combination of camera-based recognition and sensor data to enhance the system’s responsiveness and reliability.

\subsection{Haptic feedback} in gesture recognition systems provides tactile responses to user interactions, enhancing the immersive experience of virtual environments. It simulates the touch and feel of virtual objects, providing critical feedback that makes the interaction feel more real and tangible. This can be achieved through various mechanisms such as vibration motors, force feedback devices, or wearable haptic suits. The sophistication of haptic feedback varies widely, from simple vibrations indicating a successful gesture recognition to complex force feedback that replicates the texture, resistance, and weight of virtual objects.

\subsection{Industrial design}
The industrial design of hand gesture recognition systems focuses on the physical form factor of the devices used to capture and interpret gestures. Ergonomics plays a significant role in the design process, ensuring that devices are comfortable for prolonged use without causing user fatigue. This includes considerations for device weight, adjustability for different hand sizes, and non-intrusive designs that do not hinder natural movement. The aesthetics of the device also need to be appealing to users, potentially influencing the adoption rate of the technology.

\subsection{Discussion}
The integration of spatial positioning, gesture tracking, haptic feedback, and thoughtful industrial design are crucial for creating effective and user-friendly systems.

Spatial positioning and gesture tracking are foundational for any gesture-based interaction system. The accuracy of these systems affects not only the user's ability to control the environment effectively but also impacts the system’s ability to offer real-time responses, which are crucial for applications requiring high levels of interaction such as virtual reality gaming or surgical simulations. The choice of technology—whether infrared, ultrasonic, camera-based, or a hybrid approach—needs to align with the specific requirements of the application, such as the need for precision, the environmental conditions under which the system will operate, and the computational resources available.

Haptic feedback is another significant aspect that enhances the immersive experience and user satisfaction. The challenge lies in seamlessly integrating feedback mechanisms that are responsive enough to provide real-time interaction cues without overwhelming the user or requiring too much computational power. Innovations in materials science and actuator technology could play pivotal roles in developing new haptic devices that are more effective and less cumbersome.

Industrial design focuses on user comfort and system durability, impacting the long-term viability and user adoption of technology. A well-designed gesture recognition system must be adaptable to various user environments and needs to consider factors like device wearability, ease of use, and maintenance. As gesture-based interactions become more common in consumer electronics, the aesthetic aspects of these devices will also become increasingly important.

Finally, the successful deployment of gesture recognition technologies depends on a careful balance between technical performance and user-centric design principles. Future research should focus on addressing the limitations of current technologies, such as improving the accuracy of gesture recognition in complex backgrounds or developing more sophisticated algorithms that can learn and adapt to individual user preferences. The evolution of artificial intelligence and machine learning will undoubtedly play a crucial role in achieving these improvements. Integrating these technologies into gesture recognition systems will facilitate more natural and intuitive user interfaces, potentially transforming how we interact with digital devices and environments.

\section{Importance of the identified factors to gesture input hardware selection}
\label{sec: Holistic importance}
There are advantages and disadvantages to each of the four gesture data capture schemes mentioned above based on the four objectives listed above. The computer vision solution cannot solve the reliability problem of hand gesture tracking. The constant need to face the camera can easily cause user fatigue and cannot provide haptic feedback. Data can only be obtained in a limited environment, mainly relying on algorithms and a large amount of training data. For the EMG control scheme, it cannot achieve spatial positioning and gesture tracking, and tactile feedback could not solve well enough because the electromyographic signal mainly reflects muscle contraction, generates bioelectricity, and individual muscle physiological changes a lot, so it is difficult to achieve a universal model. Still, relative to other gesture recognition, the myoelectricity scheme has many features: The first is the low cost because it is based on algorithms with no particular components requirements; The second is low power consumption. The power consumption of products sold in the current market is less than 0.1W, and continuous operation can exceed ten h. Third, it is an active technology with a high degree of freedom. The outside world does not affect it; an interactive experience is natural. It can provide a scheme combining gesture and posture. Solution for data gloves, space location, and gesture tracking has been implemented with better ability. It can be realized through the design of different sensors to different tactile feedback; compared to the visual design, the scheme has little influence on the environment, a high recognition rate, strong real-time etc. properties, which is the direction of the more promising, but now the main disadvantage is that the cost is too high, according to the analysis of industrial design, how to design a cheap, lightweight, high recognition accuracy, low delay data glove is a research direction in the future. For radar solutions, the primary use is to obtain doppler frequency signal data. Compared with the data glove, its spatial position and gesture tracking ability are weaker because beyond a certain distance of the receiving antenna, signal interference by the environment quickly, as it cannot provide tactile feedback. Still, compared with the data glove, it has the advantage of low cost.

Although the advantages and disadvantages of each scheme are different, each method has its specific application scenarios for computer vision solutions, usually used in the process of application scenarios like live video or photos, combined with the user's gestures (e.g., thumb up, finger heart), real-time increase corresponding stickers or special effects, rich interaction experience; Intelligent home, intelligent household appliances, household robots and other hardware devices, through the user gesture control corresponding functions, human-computer interaction more thoughtful, natural; Intelligent driving, gesture recognition could be applied to the driving assistance system, using gestures to control various processes and parameters in the car, to a certain extent, emancipate user's eyes, put more attention on the road, and improve driving safety. For the data glove solution, common application scenarios like 3D modelling design, the use of data gloves can enable designers to have more immersive interaction on the 3D design platform to ensure that the design of products can reach a high degree of accuracy; VR games, through data gloves can enhance the user's sense of immersive experience. For EMG solutions, the main application scenarios are in medical rehabilitation and sports fitness. When users are not suitable for large-scale gestures, EMG bracelets can assist users in achieving corresponding goals. The radar solution can be applied to intelligent homes, speakers, and other Internet of Things (IoT) devices. For example, suppose the user stands in the middle of the bedroom and waves his hand. In that case, the curtain will automatically close/open, or the user is in the study aside to knock in the air continuously two times. The soundbox will automatically switch the background music, providing a more suitable office environment for the user.


\section{Future Research agendas}\label{sec: Future Research agendas}

Although various experimental studies have been carried out in gesture recognition, there are still many problems and research challenges to be solved in this field \cite{al-shamayleh_systematic_2018}.The following are several research directions and their research Motivation.

\begin{itemize}
\item How to effectively filter noise in input data, such as meaningless gesture data.
\item How to design a lightweight gesture recognition network.
\item How to design and program vibration motor to give users tactile feedback.
\item If it is possible to improve recognition accuracy by combination of eye data and gesture data.
\end{itemize}

\subsection{Filter noise in input data}
Now there is more and more research on the user experience of immersive interaction in virtual reality, so the investigation of gesture recognition technology is becoming more and more critical in this field. The first research direction that can enhance the sense of immersive interactive experience is how to eliminate the finger jitter unconsciously generated by users, which also called noise is easy to be recorded by sensors and transmitted to the input of the classification model, thus affecting the recognition accuracy of the model. Although existing methods can restore hand posture well, the problem of jitter during hand movement has not been completely solved. Jitter is an involuntary movement accompanied by an expected gesture or hand movement that causes the hand posture to deviate from the user's intention \cite{leng2021stable}. However, it is difficult to improve this problem under the conditions of some hardware devices, such as a single camera based on visual recognition technology, because a single camera is difficult to accurately capture the changes of fingers in all directions. When hands are crossed, the camera cannot accurately obtain the real-time changes of the covered hands. There are also EMG bracelets based on EMG signals, which are mainly used in medical rehabilitation. However, immersive virtual interaction should conform to users' daily behaviours to enhance users' actual experiences. Therefore, EMG bracelets are not suitable for research in this direction. The other is radar-based equipment, which is easy to be interfered with by the environment. So the best gesture recognition equipment for the study is the data glove, which can accurately capture the gestures in different directions, will not be influenced by the external environment. Even though it can also be spread by accelerometer and gyroscope to calculate the angle change in the hand crossed state. Hence, the data glove is more suitable for the direction of research. The problem is that the user is biologically unconscious. Hence, it is challenging to train the user to eliminate the effect, so it could be ameliorated by hardware or software models. For the data glove as the data acquisition device, gesture data in data glove is the main contributions of the gyroscope and accelerometer, therefore, how to filter the noise data can be first determines these sensors whether high frequency or low frequency data were collected. For accelerometer in normal gestures change should be in a state of low frequency, When the acceleration value of high frequency is detected, noise is usually generated. However, gyroscope should be in high frequency state in normal state, and it often produces low frequency drift due to the loss of reference frame. Therefore, when detecting low frequency value, it can be considered to filter out the data.So according to these two conditions we can use different filtering methods to carry out experiments. 

\noindent \textbf{Scale-invariant feature transform(SIFT)}
In 2017, Leite \textit{et al.} \cite{leite2017hand} presented a real-time framework that combines depth data with infrared laser speckle pattern (ILSP) images captured from Kinect devices for static gesture recognition and interaction with CAVE applications. Background removal and hand position detection on system startup is performed using only depth maps. After that, start tracking with the hand position of the previous frame to find the hand centroid of the current structure. The obtained points are used as seeds of the region growth algorithm and are hand-segmented in the depth map. The result is a mask used for hand-split sequences of ILSP frames. Also, they \cite{leite2017hand} applied action restrictions for gesture positioning so that each image is marked as ``gesture'' or ``non-gesture''. The ILSP corresponding frames marked as ``gesture'' are enhanced by mask subtraction, contrast stretching, median filtering and histogram equalization. The results as input were extracted using a scale-invariant feature transform (SIFT) algorithm, visual word bag construction, and multi-class support vector machine (SVM) classifier. Finally, they \cite{leite2017hand} set up a syntax based on gesture classes to translate the classification results in control commands into a CAVE application. Tests and comparisons were performed to show that the implemented plug-in is an effective solution. We achieved state-of-the-art recognition accuracy and efficient object manipulation in a virtual scene visualized in a CAVE.

\noindent \textbf{Particle filter(PF)}
Gesture recognition has been widely used in intelligent sensing, robot control and intelligent guidance. In 2018, Wang \textit{et al.} \cite{wang2018effective} presented a gesture recognition method based on the inertial sensor. This method uses a position estimator to obtain the trajectory of hand motion. The technique uses attitude estimation to generate velocity and position estimation. The attitude quaternion of gyroscopes, accelerometers and magnetometers is estimated by a particle filter (PF). The improved method is based on the resampling method to make the original filter converge faster. After smoothing, the track is converted into a low-definition image and then sent to the recognizer based on BP-NN (BackPropagation Neural Network) for matching. Experiments on real hardware prove the effectiveness and uniqueness of this method. Compared with the typical accelerometer or vision sensor, this method is fast, reliable and accurate.

\noindent \textbf{Low-pass filter(LPF)}
Mummadi \textit{et al.} proposed a sign language recognition system based on data gloves \cite{mummadi2017real}. The method uses a low-pass filter to filter the high-frequency noise generated by the accelerometer and a high-pass filter to filter the low-frequency drift noise generated by the gyroscope. Finally, the accelerometer and the gyroscope data are fused as the input of the KNN model. Through this method, the model's input data can be smoother, and most of the noise data can be filtered. Finally, the sample data is collected from 57 participants, and the average accuracy is 92\%, and the delay is lower than 65ms. Cheng \textit{et al.} \cite{cheng2020anti} proposed an anti-interference method, designed a state feedback controller and studied the controller design for uncertain systems with input disturbances.

\noindent \textbf{Finite state machine(FSM)}
A novel FSM algorithm based on finger gesture tracking is proposed by \cite{hakim2020enhanced}. The method consists of two main algorithms: one is to build FSM- Builder system to extract FSM models from finite repetitive action sequences; The system aggregates similar hand positions into one and represents them as one state. Each state defines the correct finger position and its relationship to other states. This relationship is then used in the second part of the FSM-FT system, called the FSM-Runner system. Given the FSM model of FSM-Builder, the Runner system helps guide the input hand image and point cloud into the correct state of the hand, resulting in the correct finger position. Experimental results show that the recognition rate of the system is 83\% on average \cite{hakim2020enhanced}. However, in some states, the accuracy of the results is not high because the selected features are not sufficient to distinguish each category. Another reason is insufficient variation in training data due to the difficulty of collecting training data for certain specific types. Future studies can apply deep learning techniques and convolutional neural networks such as FSM-FT Runner to extract two-dimensional features.

\subsection{Design a lightweight network}
The number of network parameters is easy to affect the response speed of the network, because the more parameters need to occupy more computing resources, so how to design a lightweight network is an important part of gesture recognition. In terms of simplified network, structural pruning is mainly considered, including three main methods such as sparse connection, tensor decomposition and channel pruning.

\noindent \textbf{Structure pruning}
High computational complexity and frequent memory accesses make real-time deep learning algorithms difficult to implement. Network pruning is a promising method of solving this problem. Unfortunately, pruning results in irregular network connections that require additional representation efforts and do not fit well with parallel computation. Anwar \textit{et al.} \cite{anwar2017structured} introduced structured sparsity at various scales for convolutional neural networks: feature map-wise, kernel-wise, and intra-kernel strided sparsity. This structured sparsity is advantageous for direct computational resource savings on embedded computers, parallel computing environments, and hardware-based systems.\newline{}

\noindent \textbf{Sparse connection}:
There are a lot of connections in a network, and the basic idea is to get rid of unimportant connections and make them sparse. Although it can reduce the model size of the network, it may not reduce the running time of the network. Deep neural networks have achieved remarkable performance in image classification and target detection, but the cost is a large number of parameters and computational complexity. So Liu \textit{et al.} \cite{liu2015sparse} showed how to use sparse decomposition to reduce redundancy in these parameters. By using inter-channel and intra-channel redundancy, the identification loss caused by maximising sparsity is minimised through fine-tuning steps to obtain the maximum sparsity. At the same time, they also proposed an efficient sparse matrix multiplication algorithm based on CPU for sparse convolutional neural network model \cite{liu2015sparse}. Their CPU implementation proved to be more efficient than existing sparse matrix libraries, achieving significant acceleration over the original dense network.\newline{}

\noindent \textbf{Tensor decomposition} 
Tensor decomposition is to take a parameter matrix of convolution network through tensor decomposition and use its low-rank property to approximate it. Data loss is inevitable when collecting traffic data from an intelligent transportation system. Previous studies have shown the method's advantages based on tensor completion in solving multidimensional data filling problems. Chen \textit{et al.} \cite{chen2019bayesian} extend the Bayesian probabilistic matrix decomposition model proposed by Salakhutdinov and Mnih(2008) to higher-order tensors and apply it to Spatio-temporal traffic data input tasks. During experiments, they not only care about the model configuration but also are concerned with the representation of data (that is, the matrix, the third-order tensor and fourth-order tensor) and evaluate the performance of the fully Bayesian model, and through a large number of experiments to explore different ways of data representation affect how the imputation performance experiments also show that the data representation is a key factor model of performance, third-order tensor structures are superior to matrices and fourth-order tensor representations in preserving data set information.

\noindent \textbf{Channel pruning} 
Channel pruning is to remove some channels simply and rudely after training a network. He \textit{et al.} \cite{he2017channel} introduced a new channel pruning method to accelerate the deep convolution neural network. Given a trained CNN model, and proposed a two-step iterative algorithm to efficiently prune each layer through channel selection and least square reconstruction based on LASSO regression. The algorithm is extended to multi-layer and multi-branch cases. This approach reduces cumulative error and enhances compatibility with various architectures. Their pruned VGG-16 achieved five times the acceleration of the state-of-the-art results, increasing the error by just 0.3 per cent. More importantly, this method can speed up modern networks such as ResNet and Xception, and the accuracy loss is only 1.4\% and 1.0\%, respectively, under 2x acceleration, which is very significant.

A feasible solution is to gradually reduce the depth and width of the model for the model that has been trained in advance under the premise of ensuring accuracy. When the depth of the model is reduced to a certain number of layers, there will be a critical point where the depth is no longer reduced, and the width of the model is reduced. Similarly, when the width is reduced to a critical point, it indicates that the network has been scaled to the minimum structure with acceptable accuracy. So, we can use this method to slim down the model.

\subsection{Design tactile feedback}
Haptic feedback has always been considered as an important step towards achieving full VR immersion, and haptic feedback in VR is mostly achieved through the application of haptic gloves or handheld devices to give a sense of touch. At present, visual and auditory sensory stimulation has been greatly developed in VR scenes. If tactile stimulation can be achieved, users will enjoy a more realistic and immersive VR experience, because the effective touch feeling in virtual environment can make users more integrated into the environment. Due to network delay, users may need to wait for a period to receive feedback after performing certain operations. If the time is too long, the user may suspect that he or she is operating improperly. But if we add vibration feedback, it will reassure the user before other feedback comes along.

In 2018, Shigapov \textit{et al.} \cite{shigapov2018design} introduced a kind of specially designed virtual reality system of low-cost digital sensor gloves, gloves with the existing concept on the function and design is different, they have a variety of functions, including feedback, tactile feedback of discharge, the feeling of bending fingers, fingers grip strength and motion prediction and three-dimensional positioning, in order to improve the sense and practice experience of virtual reality. Manual dynamic perception and freedom of movement, common in the real world, provide instant information about objects in the virtual world. In VR, digital gloves act as remote controls and provide physical feedback when users touch virtual objects. The glove allows proximal and distal finger joint movement and position/orientation determination by an inertial measurement unit. These sensors and haptic feedback caused by the vibration patterns of fingertip coins are integrated into a wireless, easy-to-use and open-source system.

Augmented reality (AR) technology has developed rapidly in the past few years and is being used in different fields. One of the successful applications of AR is immersive and interactive serious games that can be used for educational and learning purposes. A prototype AR serious game has been developed and demonstrated \cite{zhu2020development} . Players explore and interact with the environment using a headset and vibrating haptic feedback jacket in this augmented reality game. Fourteen vibration actuators are embedded in the vibrotactile feedback jacket to create an immersive AR experience. These vibrators are triggered according to the script of the game. With 14 vibration actuators placed in different body positions, the tactile feedback jacket can generate different vibration patterns and intensity levels to provide real-world feedback\cite{zhu2020development}. There are different combinations of vibration modes and intensities in each game scenario. The performance of the AR serious game prototype is evaluated and analysed. The computational load and resource utilization of normal and heavy game scenarios are compared.

\subsection{Combination of eye data and gesture data}
Eye tracking recognition and gesture tracking recognition are complementary technologies in virtual reality. At present, most researchers focus on single eye tracking recognition or single gesture tracking recognition. However, in the virtual reality environment, eye tracking and gesture tracking and positioning are related to some extent. Just as in the real world, people produce a behaviour through hand-eye collaboration, so the immersive experience in virtual reality could also be achieved through hand-eye collaboration. Therefore, the influence of hand-eye collaboration on gestures or eye-tracking recognition can be further studied.

With the function of the user and the vehicle's gesture interaction system, Roider \textit{et al.} \cite{roider2018see} put forward an integrated gaze data and algorithm to improve the accuracy of pointing gesture data. The experiment found that it can improve the pointing accuracy of fingers. Still, the benefit depends on user gesture of initial state and the position of the target element; the results show that the direction of improving accuracy gestures of great potential. Besides, supporting natural communication is fundamental to people working remotely and face to face. In 2020, Bai \textit{et al.} \cite{bai2020user} investigated how remote experts shared enhanced gaze and gestures with local users to improve the user experience. They found that integrating gaze and gestures performed better than using gaze or gestures alone and scored significantly better by combining the two features than using gestures or gaze alone. With the development of sensor technology, in 2021, Schneider \textit{et al.} \cite{schneider2021gesture} used modal sensors to study binary interaction of empirical research and theoretical framework, focused mainly on the perception of gaze and gestures, through the study found that the current research status of the field or in the early, and discussed the future development that it needs more complex algorithms to support this idea.

\section{Discussion} 
\label{sec: Discussion}
\begin{figure}[h]
  \centering
  \includegraphics[width=8cm]{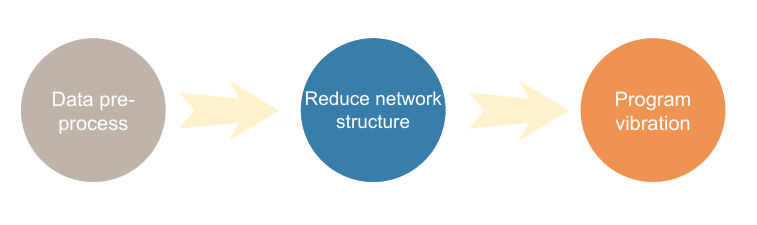}
  \caption{development process of hand gesture recognition system} \label{fig:process}
\end{figure}

According to the potential research directions mentioned earlier, we can design a future gesture recognition system development process, as shown in the Fig.~\ref{fig:process}, first collect data and the data preprocessing, and then design a network model and optimize the model (by simplifying network technology), then add tactile feedback of gesture recognition system settings (through vibration feedback programming). According to this development process, we can establish the whole gesture recognition system step by step, from data collection to pre-processing to the final tactile design; each step is the foundation of the following process. Only by doing the preliminary work well can we achieve the following results.

However, we need to focus more on data processing in the development process. As Andrew Ng advocate data-centric AI, because in the past decade development of neural network model, neural network architecture is already very mature, so the fixed neural network architecture, find out the ways to improve data will be more efficient. As for small data, Andrew NG believes it can also be powerful: ``Just having 50 good examples is enough to explain to a neural network what we want it to learn."

\section{Conclusion} \label{sec: Conclusion}
This paper summarises mainstream gesture recognition equipment and discusses its advantages and disadvantages and different application scenarios. At the same time, we summarise the common data set of gesture recognition and introduce the quaternion data expression and the way of data pre-process Principal Component Analysis (PCA). Moreover, the neural network models commonly used in gesture recognition are introduced, and their characteristics are described. Finally, several potential research directions are summarised through the literature analysis, and according to these research directions, a future development framework of the gesture recognition system is proposed, which is mainly composed of three parts. The first part is data preprocessing, which is used for data filtering and dimensionality reduction. It can effectively reduce the size and quantity of input data of the classification model and reduce the working pressure of the model. The second part is the model optimization part. For now, a better solution is to optimize the depth and width of the model to meet the requirements of lightweight models. The third part is to increase the tactile feedback to enhance the user's sense of immersion. The last part further discusses the relationship between eyeball and gesture through the analysis of the previous parts. Future research is not limited to the above parts. For example, we can add interpretable artificial intelligence to help users better understand operations in virtual scenarios.


\bibliographystyle{unsrt}  
\bibliography{references}

\end{document}